# ONLINE MEASUREMENT OF THE ENERGY SPREAD OF MULTI-TURN BEAM IN THE FERMILAB BOOSTER AT INJECTION*

J. Nelson, Brown University, Providence, RI, U.S. A.
C.M. Bhat† and B. S. Hendricks, Fermilab, Batavia IL, U.S.A.,

*Abstract*

We have developed a computer program interfaced with the ACNET environment of Fermilab accelerators to measure energy spread of the proton beam from the LINAC at an injection into the Booster. It uses a digitizing oscilloscope and provides users an ability to configure the scope settings for optimal data acquisition from a resistive wall current monitor. When the program is launched, a) a *one shot timeline* is generated to initiate beam injection into the Booster, b) a gap of about 40 ns is produced in the injected beam using a set of fast kickers, c) collects line charge distribution data from the wall current monitor for the first 200 μs from the injection and d) performs complete data analysis to extract full beam energy spread of the beam. The program also gives the option to store the data for offline analyses. We illustrate a case with an example. We also present results on beam energy spread as a function of beam intensity from recent measurements.

## OVERVIEW

In recent years Fermilab has undertaken significant improvements to the existing accelerators to meet its high intensity proton demands for accelerator based HEP experiments onsite as well as long baseline neutrino experiments. One of the important aspects of this program is "Proton Improvement Plan" [1] with a baseline goal to extract the beam at 15 Hz rate from the Booster all the time with about 4.6E12 p/Booster cycle. With PIP -II [2], it is foreseen to increase the Booster beam delivery cycle rate from 15 Hz to 20 Hz, replace the existing normal conducting LINAC by superconducting RF LINAC and increase beam intensity by 50%. Hence, the current Booster plays a very significant role at least the next one and half decades in the Fermilab HEP program.

In support of the proposed upgrades to the Booster, a thorough understanding of the properties of the beam at injection is crucial; in particular, monitoring energy spread of the beam from the LINAC after the completion of injection. In the past, many attempts have been made to measure beam energy spread (e.g., ref. [3]) in bits and pieces using up dedicated beam time; often such measurements are carried out only soon after a major maintenance period. Furthermore, all the past measurements were on partial turn beam in the Booster. During 2013-2014, we developed a very robust method [4] to measure the beam energy spread on the multi-turn beam after creating a notch of width ≈ 40 nsec in the injected beam. These attempts lead us to develop an application software to automate beam energy spread measurement at injection on request. In this paper we explain the general principle, various functionalities of the software and illustrate with recent measurements.

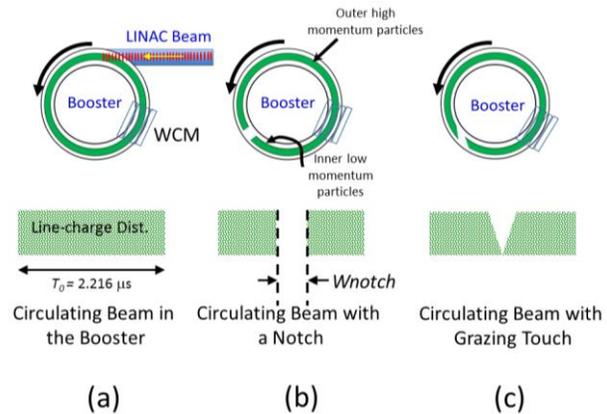

Figure 1: Schematic of newly injected multi-turn beam from the LINAC with a notch and notch filling due to slippage of the particles.

## APPLICATION SOFTWARE

*Physics Behind the Program:*

Particle beam that exits a pre-accelerator will always be having an energy spread $\Delta E$(full) about its mean (synchronous) energy $E_s$. When such a beam is injected into a synchrotron and accumulates, the higher and lower momentum particles slip differently relative to the synchronous particle while they circulate in the synchrotron. A schematic view of particle distribution of such a beam injected from the LINAC into the Booster is shown in Fig. 1(a). However, to differentiate between fast moving particles from the slow moving particles, one can create a notch (Fig. 1(b)) of width "*Wnotch*" in a fully debunched beam using a fast beam kicker and monitor the slippage of particles relative to particles with synchronous energy (see Fig. 1(c)). Evolution of line-charge distribution measured using a wall current monitor (WCM) for each case is shown in Fig 1 for clarity. By measuring the notch width and the time required for the highest momentum particles to cross (or touch in WCM data) the lowest momentum particles, "*Tgraze*", we can extract the beam energy spread using,

$$\Delta E = \frac{\beta^2 E_s}{|\eta|} \frac{Wnotch}{Tgraze}$$

Where, $\beta$ is the relativistic speed and $\eta$ is the slip factor of the Booster lattice (≈-0.4582).

___________________
* Work supported by Fermi Research Alliance, LLC under Contract No. De-AC02-07CH11359 with the United States Department of Energy
† cbhat@fnal.gov

*Program Overview and Measurement:*

The Fermilab Booster is a 15 Hz rapid cycling synchrotron built using combined function dipole magnets. The Booster receives beam from the 400 MeV LINAC with 200 MHz bunch structure and operates between an injection kinetic energy of 400 MeV and extraction kinetic energy of 8000 MeV for protons. The beam revolution period at injection in the Booster is $T_0 \approx 2.216$ μs. A single beam pulse from the LINAC is up to about 60 μs, capable of providing > 25 Booster turn beam at a beam delivery rate of 25 mA. However, currently the maximum number of Booster turn is set to be about 18 for operational safety. Generally the 200 MHz bunches from the LINAC do exhibit variation in the intensity and possibly a small variation in the beam energy spread through the pulse. Figure 2 depicts typical LINAC pulses for 4, 13, 14 Booster turn beam. Hence, after the completion of the injection it may be required to allow a few revolutions to settle down and to give a constant line charge density DC beam before the notch is produced. In addition to this, there is an additional complication from bunch formation due to a small amount of leftover RF voltage from the main accelerating RF cavity of the Booster, unless the power amplifiers are turned off completely.

The application program was designed to be used in Fermilab ACNET Console Environment. A schematic view of the flow chart representing the structure of the soft-ware

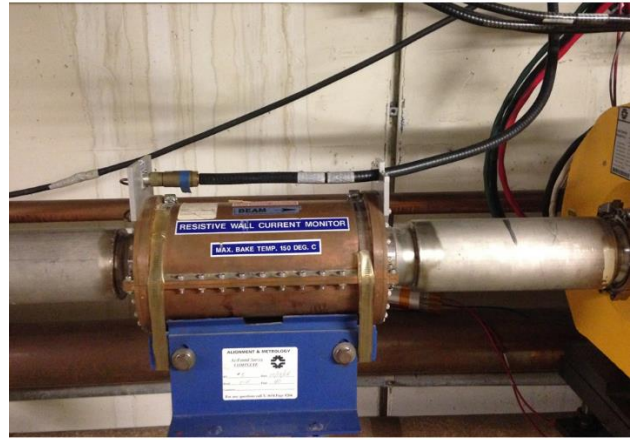

Figure 3: Wall Current Monitor in the Booster ring.

program is shown in Fig. 4. This program controls various types of hardware that are necessary to measure the energy spread of the beam. The program works as follows:

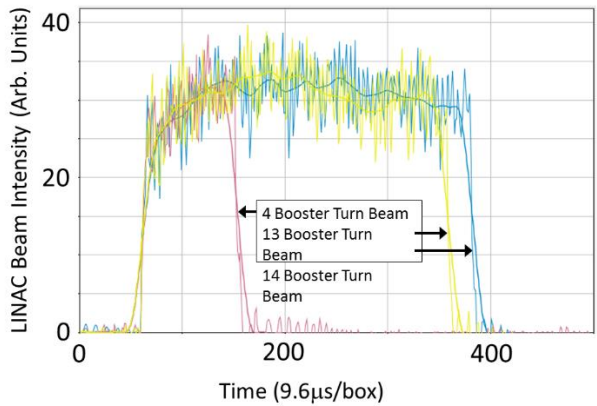

Figure 2: Typical LINAC beam pulses jest before injection into the Booster.

To measure the longitudinal properties of the circulating beam in the Booster one uses a 6 GHz bandwidth wall current monitor (WCM) [5] (see Figure 3). A fully debunched beam cannot be detected using the WCM; for such a beam the WCM signal looks like the one shown in Figure 1(a). As soon as a notch is produced using a set of fast kickers [6], one can see distinctly the region without beam and one with beam; an integration of the WCM signal for period $T_0$ scales as the beam intensity. By taking the WCM data for several revolutions one can study the time evolution of the notch. A Tektronix, TDS7154B Digital Phosphor Oscilloscope of type 1.5GHz 20GS/s is used in the Booster to record the WCM data.

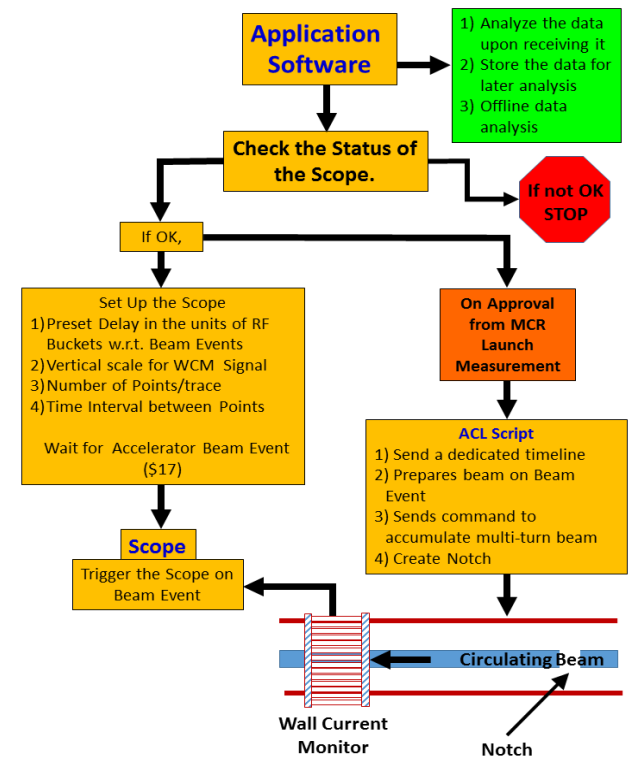

Figure 4: The flow chart of the application software.

1) Check if the executing the program is safe (all necessary instruments are available to measure the Booster energy spread) and warns if another user is using the same devices. If so, one cannot proceed.

2) If it is ok, then user can send scope set up. Then the program sets necessary parameters on the Tektronix scope *viz.*, a) a window for receiving WCM data with upper and lower amplitude levels (default values are 1 V and -.04 V, respectively, if WCM signal saturates than user need to set properly), b) number of data points (default is 50000 for "Record length"), c) space between the data points ("interval" 0.4 ns), d) beam event of interest and e) a constant delay with respect to beam event to start collecting the data.
3) If the scope is available, then program executes another software written in *accelerator command language* (ACL) upon permission from the Main Control Room (MCR) crew chief which enables a) a "*1-shot timeline*. (A timeline is a user requested sequence of different types of accelerator system *clock* events, with and without beam, put in certain order to make accelerator operation smooth. This is specific to Fermilab ACNET environment), b) moves all timers related to RF paraphrase, beam feedback to later settings predetermined by the user, c) moves notch timer to soon after the injection. Number of Booster turns is set ahead of issuing the ACL script. Once the ACL command is issued the rest of the beam operation is automatic.
4) Scope collects the WCM data.
5) Analysis the data to extract the beam energy spread.
6) Gives a graphical output on a separate window for the user's perusal.

Some highlights of this programs functionality is, it is the first application program developed for the Fermilab Booster that can measure the beam energy spread of a multi-turn beam in the Booster at injection. This program gives an ability to measure the beam energy spread with minimum interruption on the operation and would help to improve the overall Booster performance.

*Analysis of the Data:*

The online analysis of the WCM data is one of the critical task of this application program. Once the data is collected the data can be can be analysed online or offline after saving it. As far as this software is concerned the analysis of the data online or offline is identical. In both cases, the program has to make several decisions. For example, a) search for the first notch and the subsequent appearance of the notch on each revolution, b) determine notch width (95%), c) time required for grazing touch and d) average revolution period, etc.,

Before developing this application program two packages, one using MATLAB environment and another using Python platform [7] were developed for off-line data analysis. These two packages helped us to benchmark the newly developed program and make it very robust.

A. *Search for the 1st Notch*

An example of the data received from the WCM for ≈100 revolution in the Booster around injection is shown in Fig. 5. The icicle type structure develops soon after the first notch is produced in the beam and as the same notch arrives at the WCM once per revolution. Electronics used in the WCM before the data arrives at the scope is a DC coupled system. Consequently, as the multi-turn beam is injected into the Booster the beam amplitude increases initially above the background level and then slowly starts to decline as beam settles down. Hence one sees formation of a structure that resembles horse *mandible* and *cervical vertebrae* shape to the line charge distribution. The increase in the amplitudes of WCM data after about 150 μs is due to increasing RF voltage.

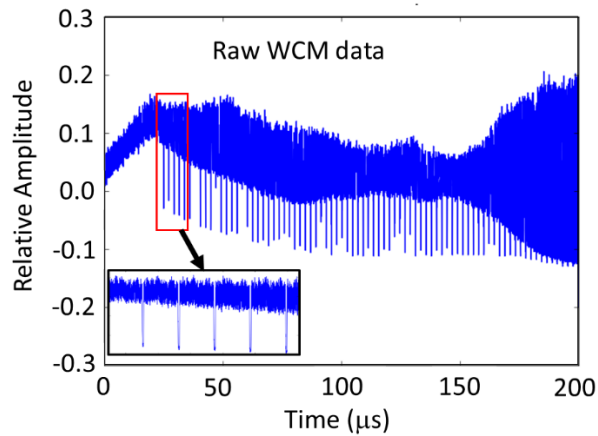

Figure 5: WCM data for the first 200 microsecond around injection. Zooming in it is possible to see a few of these notches, each is about 2.2 microseconds from each other (as shown in inset).

Over time the notch depth starts to decrease as the particles with different momenta drift toward its center and the notch begins to fade away. Until the grazing touch (for example see, picture "c" in Fig. 1), the notch depth or integrated area between two successive notches in the WCM data remains constant. Beyond the point of the grazing touch, the depth of the notch start decreasing. Therefore, by following the evolution of the notch and measuring the integrated area of WCM data for every revolution we should be able to measure the time required for grazing touch.

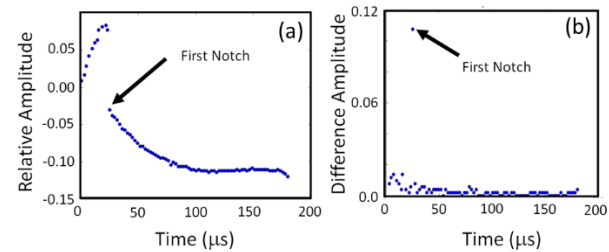

Figure 6 (a) Plot of all the minimum for every 2.2115μs. The first notch appears immediately after a large jump. (b) The absolute difference of the next point and its "left neighbour".

First the program scans the entire WCM profile shown in Fig. 5 and then it uses a guess period of 2.2115 μs to make "approximate cuts". The program then searches for the minimum in each piece and saves those amplitude values and their corresponding time coordinate. Since we are interested in knowing when the notch appeared in the WCM data for the first time and in knowing its width accurately, the program compares all minima and looks for where a sudden large jump occurs in the amplitude of the minimum. The time associated with first large excursion point in the searched set of amplitudes of minima in WCM data represents occurrence of the first notch. The program infers that the subsequent minimum from the first notch are the notch at subsequent revolutions. Figure 6 depicts result of the algorithm for a typical data set where the location of the notch is identified. To make identification of the first appearance of the notch more robust we take differential in minimum amplitude between cuts as shown in Fig. 6(b). This gives unambiguous time stamp for the first notch.

notch (turn number). The revolution period is calculated taking an average of the first few points from Fig.7. Using newly calculated revolution period we can obtain better cuts on the data. The large scatter observed in the measured revolution period arises from the uncertainty in the location of the minimum in the notch region.

Figure 8 shows the integrated area between two successive minima established as in Fig. 6 (a) as a function of turn number. The base line for each revolution is calculated as an average of the minimum amplitude at notch region. Knowing the calibration factor between image current and the signal area as measured by the scope one can measure the beam intensity at injection soon after the notch is formed.

Data in Fig. 8 show that after the grazing touch the area starts falling down. This happens because the background level at the notch starts raising up due to the slipping off momentum particles in the ring. Measuring the number of revolutions required for the grazing touch is crucial. The program makes a decision based on an algorithm. It first calculates the approximate area of the $T_{graze}$ (which is the "Top notch"-("Top notch"-"Bottom notch")*0.15) and then starting from the lowest intensity point (see Fig. 8) comparing each point to this approximate area of the Tgraze. When it reaches a value greater than or equal to this approximated value the program has found the inflection point and stops. The corresponding turn number multiplied by revolution period gives $T_{graze}$.

*Finding the notch width*

The shape of the notch even when it is formed will not be a nice rectangular well; in reality, its shape depends on the kicker pulse shape that produces the notch. Furthermore, non-zero voltage on the accelerating RF cavities, Vrf , gives rise to some level of bunching as shown in Fig. 9. (As long as the bucket height due to the Vrf is much smaller than beam energy spread, this method works very well).

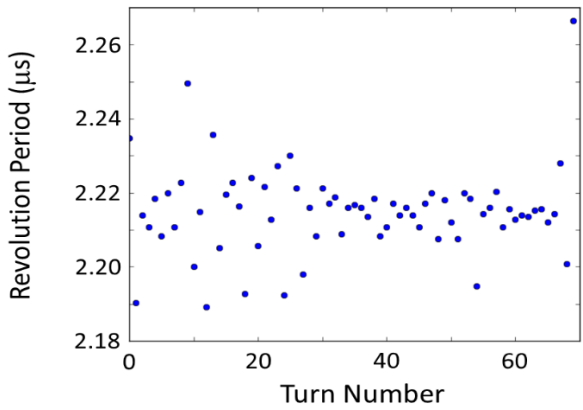

Figure 7: Revolution period versus index of appearance of the notch.

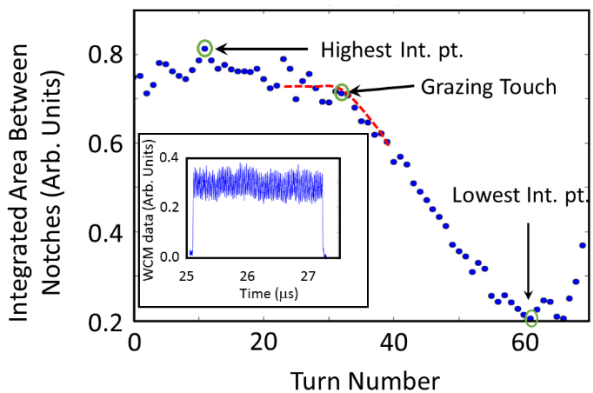

Figure 8: Revolution period versus turn number (index of appearance of the notch) from first notch. The inset shows typical beam WCM data between two successive appearances of a notch.

Figure 7 shows the time difference between the successive notch appearance and the index number of the

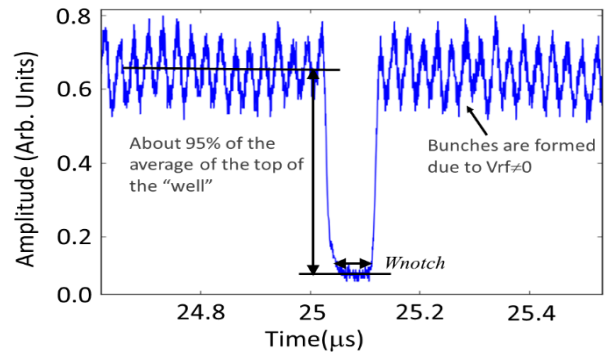

Figure 9: Width of the first notch

Consequently, the error in measured notch width will contribute a large uncertainty in the final energy spread. The average of the baseline inside the notch where there is no beam and average of the bunch signal where there is beam, were both calculated (see Fig. 9). The difference between these two averages gives the depth of the notch. Then we determine the width of the notch by searching for 95% drop on both sides of the notch. Such search is performed on the trace of the 1st notch as it appears in the WCM data.

*Data Analysis and Results*

Typical results from the data analysis are shown in Figure 10. The raw data for first 200 μs is displayed on the top. 1st notch in the beam, and an apparent beam intensity used to determine the grazing touch are shown in the middle row. The values of the measured full momentum spread "dP/P", measured number of revolutions needed for grazing touch "Graze Touch", momentum of the synchronous particles "P", the measured revolution period "Trev", and the measured notch width "NW" are displayed at the bottom for Fig. 10. The middle two figures can also be used to examine the displayed measured values qualitatively.

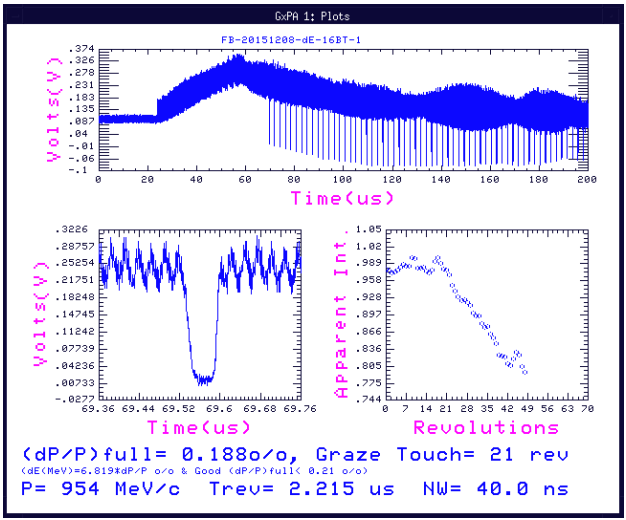

Figure 10: Result display from the program.

Figure 11 displays the measured values of the energy spread for different Booster beam intensities using the application program for the past six months. For example, the blue circles representing the measurements from March 24, 2016 for intensity in range of 1E12-6E12 p/Booster cycle did not show intensity dependence. The average energy spread from these measurement is $\Delta E$ (full) = 1.25 ±0.20 MeV.

There are multiple sources of uncertainties in the measured energy spread arising from systematics (which are not taken into account in the quoted values above)−1) the uncertainty in slip factor $\eta$ would introduce about 1%, 2) the uncertainty in the beam energy give about 0.5%, 3) error in decision on *Tgraze* expected to be about 10% and 4) error on *Wnotch* expected to be about 5%. The errors from "3" and "4" can be reduced significantly by turning off the Vrf amplifier. These sort of measurements need dedicated

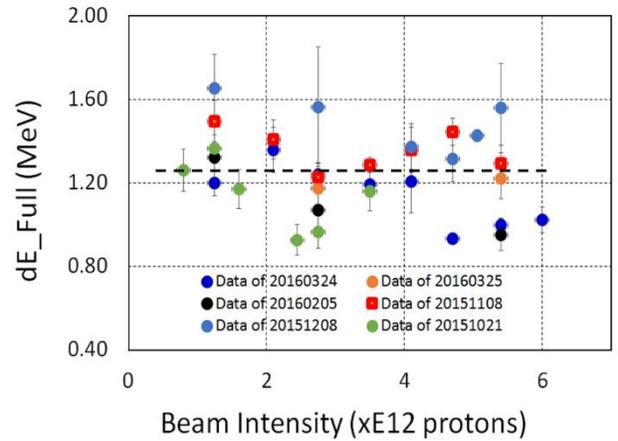

Figure 11: Measured full beam energy spread in the Booster at injection for different beam intensities over the past 7 months. The error bars are the maximum deviation from the mean values.

beam time. We expect an overall error on the measured value of $\Delta E$ (full) is about 20%.

**Acknowledgement**

Special thanks to Ming-jen Yang for many useful discussions and his help regarding data taking from the Tektronix scope. Thanks are also due to K. A. Triplett for his help in beam studies, cooperation from the MCR crew during the development of this software, and Demetrius Andrews for editing the manuscript. One of the authors J. Nelson would like to thank Fermilab 2015 SIST Summer intern program for financial support.